\UseRawInputEncoding
\documentclass[aps,prl,twocolumn,superscriptaddress]{revtex4-2}

\usepackage{graphicx}
\usepackage{xr}
\usepackage{xcolor}
\renewcommand{\v}[1]{\ensuremath{\mathbf{#1}}}
\def \AP1[#1]{\color{blue}#1\color{black}}
\def \suppsec[#1]{Supplemental Material Sec.~\ref{#1}}
\def \SF[#1]{Suppl.~Fig.~\ref{#1}}
\def \F[#1]{Fig.~\ref{#1}}

\makeatletter
\newcommand*{\addFileDependency}[1]{
  \typeout{(#1)}
  \@addtofilelist{#1}
  \IfFileExists{#1}{}{\typeout{No file #1.}}
}
\makeatother

\newcommand*{\myexternaldocument}[1]{%
    \externaldocument{#1}%
    \addFileDependency{#1.tex}%
    \addFileDependency{#1.aux}%
}

\myexternaldocument{multibandCDW_suppl_2}

\begin{document}



\title{Multiband charge density wave exposed in a transition metal dichalcogenide}


\author{\'{A}rp\'{a}d P\'{a}sztor}
\email[]{arpad.pasztor@unige.ch}
\affiliation{Department of Quantum Matter Physics, Universit\'{e} de Gen\`{e}ve, 24 quai Ernest Ansermet, CH-1211 Geneva 4, Switzerland.}
\author{Alessandro Scarfato}
\affiliation{Department of Quantum Matter Physics, Universit\'{e} de Gen\`{e}ve, 24 quai Ernest Ansermet, CH-1211 Geneva 4, Switzerland.}
\author{Marcello Spera}
\affiliation{Department of Quantum Matter Physics, Universit\'{e} de Gen\`{e}ve, 24 quai Ernest Ansermet, CH-1211 Geneva 4, Switzerland.}
\author{Felix Flicker}
\affiliation{Rudolph Peierls Centre for Theoretical Physics, University of Oxford, Department of
Physics, Clarendon Laboratory, Parks Road, Oxford OX1 3PU, United Kingdom}
\affiliation{School of Physics and Astronomy, Cardiff University, Cardiff CF24 3AA, United Kingdom}
\affiliation{School of Mathematics, University of Bristol, Bristol BS8 1TW, United Kingdom}
\author{C\'{e}line Barreteau}
\affiliation{Department of Quantum Matter Physics, Universit\'{e} de Gen\`{e}ve, 24 quai Ernest Ansermet, CH-1211 Geneva 4, Switzerland.}
\author{Enrico Giannini}
\affiliation{Department of Quantum Matter Physics, Universit\'{e} de Gen\`{e}ve, 24 quai Ernest Ansermet, CH-1211 Geneva 4, Switzerland.}
\author{Jasper van Wezel}
\affiliation{Institute for Theoretical Physics Amsterdam and Delta Institute for Theoretical Physics,
University of Amsterdam, Science Park 904, 1098 XH Amsterdam, The Netherlands}
\author{Christoph Renner}
\email[]{christoph.renner@unige.ch}
\affiliation{Department of Quantum Matter Physics, Universit\'{e} de Gen\`{e}ve, 24 quai Ernest Ansermet, CH-1211 Geneva 4, Switzerland.}





\maketitle


\textbf{In the presence of multiple bands, well-known electronic instabilities may acquire new complexity. While multiband superconductivity is the subject of extensive studies, the possibility of multiband charge density waves (CDWs) has been largely ignored so far. Here, combining energy dependent scanning tunnelling microscopy (STM) topography with a simple model of the charge modulations and a self-consistent calculation of the CDW gap, we find evidence for a multiband CDW in 2H-NbSe$_2$. This CDW not only involves the opening of a gap on the inner band around the K-point, but also on the outer band. This leads to spatially out-of-phase charge modulations from electrons on these two bands, which we detect through a characteristic energy dependence of the CDW contrast in STM images.}

Imposing a new periodicity on a crystal leads to the reorganization of the electronic bands of the parent compound through their back-folding on the new Brillouin zone. New periodicities may be engineered in designer materials, for instance in artificial heterostructures with Moir\'{e} minigaps, or emerge due to a structural or electronic phase transition. The charge density wave state is an electronic order where the charge density develops a spatial modulation concomitantly to a periodic distortion of the atomic lattice and the opening of a gap in the quasi-particle spectrum. By reducing the electronic band energy, this gap compensates for the elastic and Coulomb energy costs associated with the formation of the CDW. It also lowers the degeneracy of the electronic states at the crossings of the folded bands. These are the points in the band structure of the parent compound that are connected by the wavevector of the new periodicity. Although a gap should open at all the crossings of the folded bands, previous studies only focused on the primary CDW gap around the Fermi-level, which leads to the highest energy gain of the reconstructed system. The existence of secondary gaps and associated charge modulations remains largely unexplored.

In many cases, only a tiny fraction of the electrons are involved in the CDW formation. Therefore, the CDW gap manifests only as a slight reduction of the density of states (DOS) -which can depend on momentum- rather than a  full depletion of the DOS. This makes it challenging to measure the CDW gap using spectroscopic probes such as angle resolved photoemission spectroscopy (ARPES) and scanning tunnelling spectroscopy (STS). This is particularly true for 2H-NbSe$_2$ \cite{Wang1990,Borisenko2009,Rahn2012,Soumyanarayanan2013,Arguello2014} (hereinafter simply NbSe$_2$). However, the effect of the redistributed electrons can be readily detected in topographic STM images, even for minute changes brought upon by the opening of the CDW gap as demonstrated in the following.

\begin{figure}[htp]
\includegraphics[width=\columnwidth]{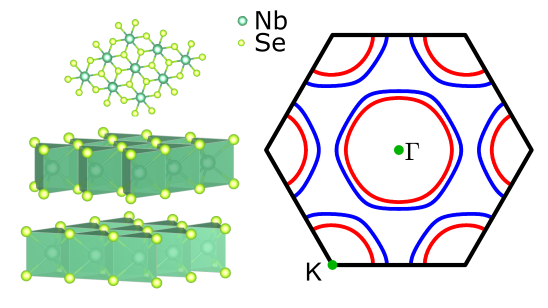}%
\caption{\label{fig:struct_FS} \textbf{Crystal structure and Fermi surface of NbSe$_2$.} Left: the three-fold symmetric crystal structure from top and side view \cite{Momma2011}. Right: the Fermi surface has been calculated using a two-band tight-binding fit to ARPES data \cite{Rahn2012}. Inner pockets (red) around $\Gamma$ and $K$ derive from one band, while the outer pockets (blue) derive from the second band; a small pancake-shaped pocket around $\Gamma$ originating mainly from Se orbitals has been omitted.}
\end{figure}

NbSe$_2$ is an iconic material of correlated electron physics. It hosts a nearly-commensurate charge density wave below $T_{CDW}=33.5$~K and a superconducting (SC) order below $T_{SC}=7.2$~K \cite{Revolinsky1965, Wilson1975, Harper1975, Long1977, Moncton1975}. NbSe$_2$ is a layered material with a three-fold symmetric crystal structure around the direction perpendicular to the layers (Fig.~\ref{fig:struct_FS}). Each unit cell is composed of two slabs of Se-Nb-Se trilayers, where the Se lattices are 60$^{\circ}$ rotated, while the Nb atoms are aligned on top of each other in a trigonal prismatic coordination with the Se atoms. 

The Fermi surface (FS) of NbSe$_2$ is mainly determined by the bonding and antibonding combinations of the Nb-4\textit{d} orbitals \cite{Johannes2006,Flicker2015natcomm} leading to double-walled barrel-shaped pockets around the K and $\Gamma$ points of the hexagonal Brillouin-zone \cite{Corcoran1994,Rossnagel2001,Borisenko2009, Rahn2012} (Fig.~\ref{fig:struct_FS}). The charge ordered state consists of three CDWs which form along the three equivalent $\Gamma M$ directions with wavevectors $(1-\delta)\frac{2}{3}|{\Gamma M}|$, where $\delta\approx0.02$ and depends on temperature \cite{Moncton1975}. In real space, this yields a locally commensurate $3a_0\times3a_0$ superstructure delimited by discommensurations \cite{McMillan1976, Pasztor2019}, where $a_0$ is the atomic periodicity.

The $3a_0\times3a_0$ reconstruction is readily accessible to topographic STM imaging. Its bias dependent contrast has been the focus of previous studies, with particular emphasis on the contrast inversion expected in a classic Peierls scenario between images acquired above and below the CDW gap \cite{Mallet1996, Sacks1998}, and on the role of defects in stabilizing the CDW~\cite{Arguello2014}. Sacks \textit{et al.} \cite{Sacks1998} calculate the bias dependence of the CDW phase in a perturbative approach, considering a single band normal state description of NbSe$_2$. They find that the phase-shift of the CDW component of the local DOS can be very different from the 180$^\circ$ expected in a one-dimensional (1D) case (\suppsec[SupSec:LDOS_1D_CDW]) when changing the imaging bias across the Fermi level ($E_F$). However, their model does not reproduce the full bias dependence of the CDW amplitude and phase that we find.

\begin{figure*}[htp!]
\includegraphics[scale=0.98]{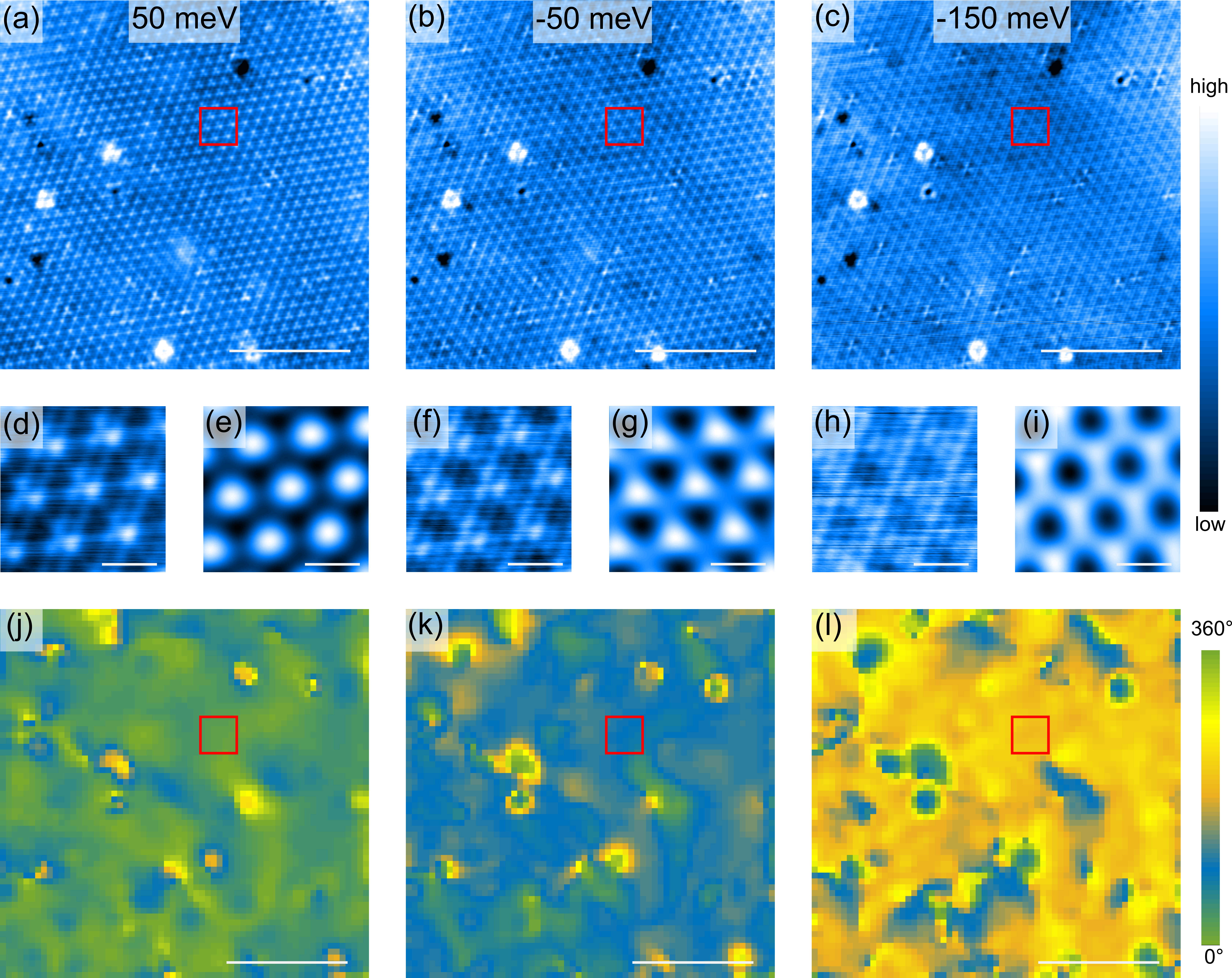}%
\caption{\label{fig:bias_dependent_topo} \textbf{Bias dependent STM imaging of NbSe$_2$ at 1.2~K.} Constant current topography showing the atomic lattice and CDW at (a) $V_{b}=50$~mV, (b) $V_{b}=-50$~mV and (c) $V_{b}=-150$~mV \cite{comment1} with $I_t=100$~pA. (d), (f) and (h) are magnified images of the areas marked by the red squares in (a),(b) and (c), respectively. The overall imaging contrast is very different in these cases, although the atomic lattice appears identical in all images (\SF[fig:suppl_atomic_lattice]). Panel (e),(g) and (i) show the magnified image of the large Fourier-filtered image of the CDWs at the same location as shown in (d),(f) and (h) respectively. It demonstrates that the variation of contrast observed in (e),(g) and (i) is stemming from a variation in the appearance of the CDW at different biases. These appearances can be quantified by a single parameter: the dephasing parameter which describes the relative position of the three CDWs.
(j)-(l) show the spatial variation of the dephasing parameter determined by fitting the CDW modulations of the STM images shown in (a)-(c), respectively. The red squares corresponds to the same area which are highlighted in (a)-(c) and magnified in (d)-(i).
Scalebar: 10 nm in (a)-(c) and (j)-(l); 1 nm in (d)-(i).}
\end{figure*}

In Figs.~\ref{fig:bias_dependent_topo}(a)-(c), we present a selection from numerous topographic STM images of the same region on a cleaved NbSe$_2$ surface at different sample biases ($V_b$) between $-0.5$~V and $0.5$~V. They show a triangular atomic lattice with a superimposed $3a_0\times3a_0$ CDW modulation (see also Suppl. Fig.~\ref{fig:suppl_atomic_lattice}(a)), consistent with previous STM studies of unstrained bulk NbSe$_2$~\cite{Mallet1996,Soumyanarayanan2013,Arguello2014,Gao2018,Castro2018nanolett,Gye2019, Guster2019,Pasztor2019}. A defect free region with a well resolved CDW outlined in red is magnified in Figs.~\ref{fig:bias_dependent_topo}(d),~(f)~and~(h) for each $V_b$. In order to identify the origin of the bias dependence of the topographic contrast in these images, we separate the atomic lattice and CDW contributions using Fourier filtering (\suppsec[SupSec:atomic_STM]). This analysis shows that the bias dependent STM contrast is due to the changing CDW signal (Fig.~\ref{fig:bias_dependent_topo}(e), (g) and (i)), since the corresponding atomic lattice contrast remains unchanged (see \SF[fig:suppl_atomic_lattice]).

 The observed CDW pattern can be modeled as the sum of three plane waves as described in Ref~\cite{Pasztor2019}. While each plane wave has its own phase $\varphi_i(\v{r})$, which depends on a selected reference point, the \textit{dephasing parameter} $\Theta(\mathbf{r})=\varphi_1(\v{r})+\varphi_2(\v{r})+\varphi_3(\v{r})~\text{mod}~360^{\circ}$ is uniquely defined for each particular CDW pattern, independent of any reference point. $\Theta(\v{r})$ represents the internal CDW structure, quantifying the local relative position of the wavefronts of the three CDWs. In Figs.~\ref{fig:bias_dependent_topo}(j)-(l), we show $\Theta(\v{r})$ corresponding to the STM images in Figs.~\ref{fig:bias_dependent_topo}(a)-(c), respectively. They were obtained by fitting the CDW contrast following the method described in Ref.~\cite{Pasztor2019}. 

Each bias voltage is characterized by a dominant dephasing parameter (Figs.~\ref{fig:bias_dependent_topo}(j)-(l)), except in the vicinity of defects discussed later. This visual assessment is confirmed by the peaked histograms of $\Theta(\v{r})$ (\SF[fig:suppl_dephasing_histogram])). Fitting a Gaussian to these histograms allows to extract a well defined dephasing parameter $\Theta_0(V_b)$ for each imaging bias (\SF[fig:suppl_dephasing_mapping]). For a quantitative analysis of the bias dependence of $\Theta_0$, we note that a given local CDW structure is represented by any arbitrary combination of $\varphi_i(\v{r})$ summing up to the same dephasing parameter, in particular the one where all three phases are equal. Moreover, the
threefold symmetry of the system implies there is no privileged plane wave among the three used to describe the CDW. These observations allow us to map the problem to a one-dimensional (1D) description with an apparent CDW phase $\varphi_0(V_b)=\Theta_0(V_b)/3$ (\suppsec[SupSec:dephasing_parameter_measured]), and model $\varphi_0(V_b)$ to understand the bias dependent CDW pattern.

\begin{figure}[htp]
\includegraphics[scale=1]{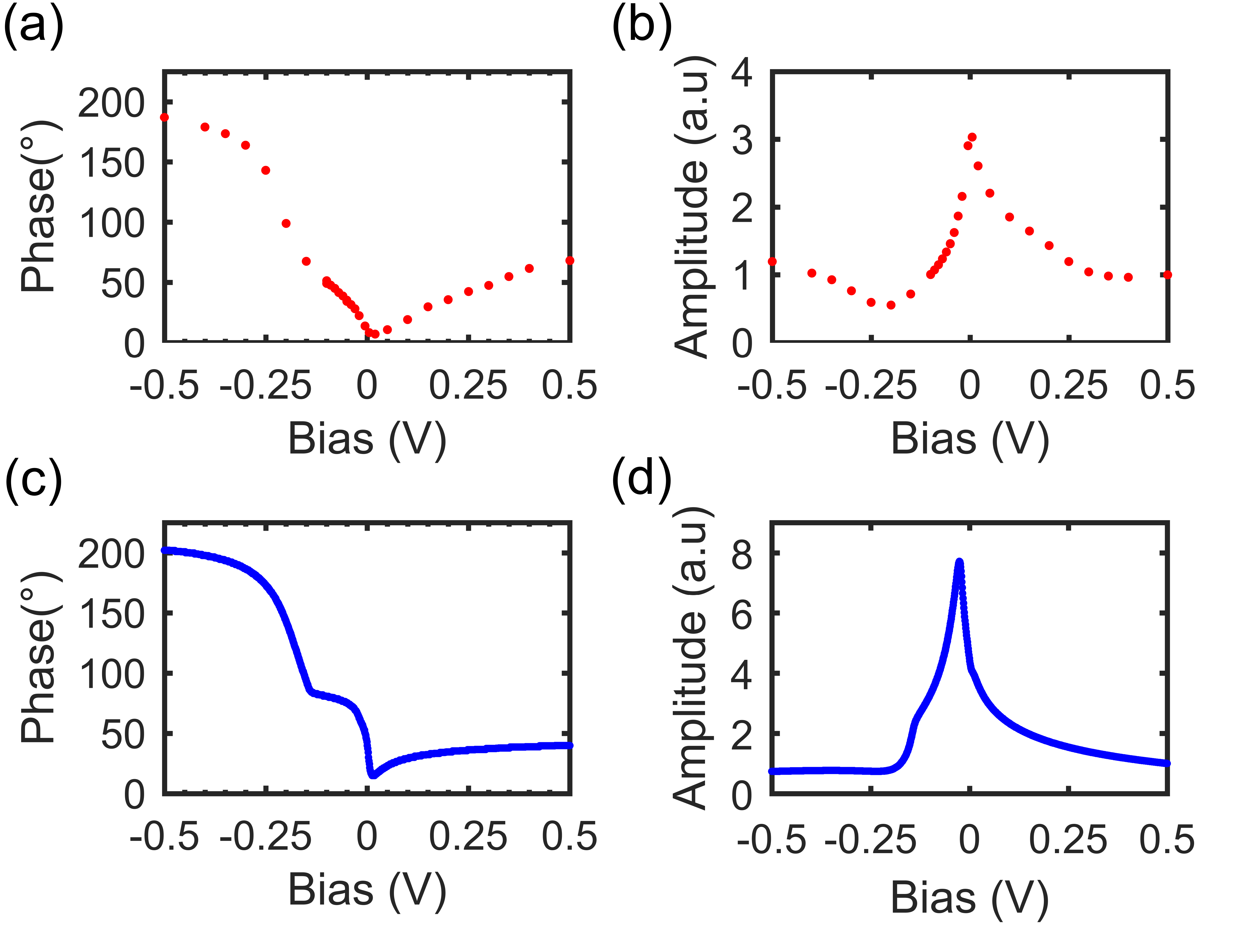}%
\caption{\label{fig:bias_dep_phase_amplitude} \textbf{Bias dependence of the phase and amplitude in experiments and in modelling.} (a) and (b) are the bias dependence of the phase and amplitude of one of the unidirectional CDW extracted from the dephasing parameter and amplitude maps. (c) and (d) the phase and amplitude of the best matching simulations in the two-gap model. The data in (b) and (d) are normalized to their $V_{b}=0.5$~V value.}
\end{figure}

Plotting $\varphi_0(V_b)$ in Fig.~\ref{fig:bias_dep_phase_amplitude}(a) reveals a striking non-monotonic bias dependence, with an inflexion point around $-0.15$~V and a minimum slightly above the Fermi-level ($E_F=0$~V). This dependence is robust as long as $\varphi_0(V_b)$ is extracted from topographic STM images away from defects (\SF[fig:suppl_away_from_defect]). Close to defects, the dephasing parameter $\Theta_0(V_b)$ is different and tends to depend much less on imaging bias (\SF[fig:suppl_around_defects]). This is consistent with earlier findings that defects (and impurities) can act as strong pinning centers \cite{Fukuyama1976,Fukuyama1978} locking the local phase of the CDW or driving the formation of CDW domains \cite{Hildebrand2016,Novello2017}. 

The CDW amplitude can be extracted in a similar way to the phase, by fitting the histogram of the local amplitudes $a_i(\v{r})$ of each plane wave measured over the entire field of view with a Gaussian, and extracting the peak value $a_i(V_b)$. The bias dependence and magnitude of $a_i(V_b)$ is nearly the same for all three CDWs (\SF[fig:suppl_amplitude](b)). For the analysis, we consider the average of these three amplitudes at each bias $A_0(V_b)=(a_1(V_b)+a_2(V_b)+a_3(V_b))/3$ plotted in Fig.~\ref{fig:bias_dep_phase_amplitude}(b).

\begin{figure*}[htp]
\includegraphics{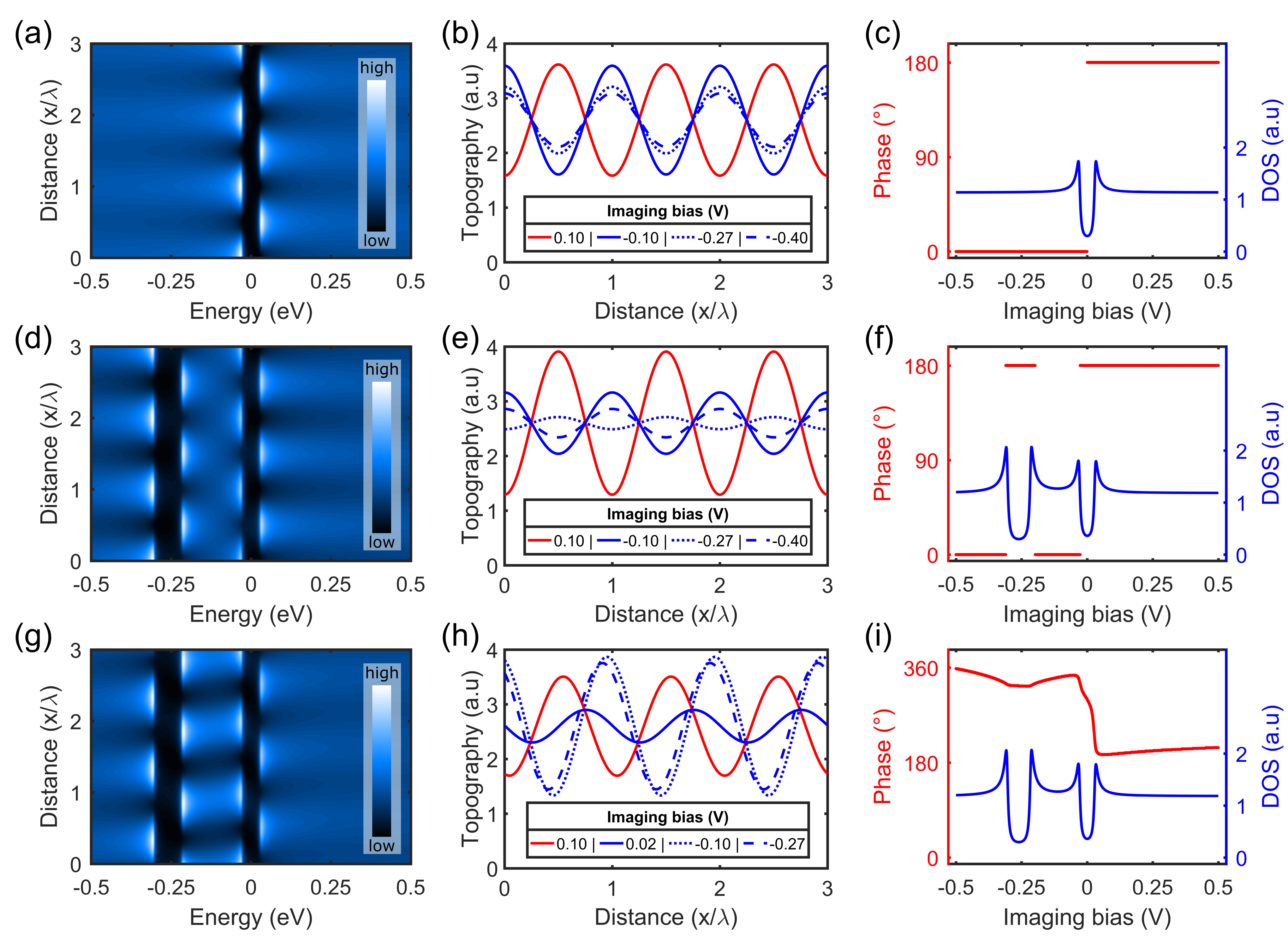}%
\caption{\label{fig:two_gap_model} \textbf{One-dimensional model description of the bias dependent phase of the CDW.} (a)-(c) There is a single CM and the corresponding gap is centred at the Fermi level.  (d)-(f) There are two CMs with two gaps which are centered at different energy (one at the Fermi-level and one below). There is no real-space phase difference between the CMs. (g)-(i) The same as (d)-(f) except that the there is a 120$^{\circ}$ (2$\pi$/3) phase difference -one atom shift- between the two CMs. In all the three cases the first column shows the spatial and energy dependent CDW LDOS maps. Second column: corresponding simulated topographic traces at selected biases. To clearly see the evolution of the phase and the amplitude the curves are vertically offset in panel (b)(e) and (h) such that they all oscillate around the same value. Third column: bias dependence of the phase (left red axis) and DOS (spatially integrated LDOS from the first column, right blue axis).  }
\end{figure*}

To understand the bias dependence of the CDW amplitude and phase in Fig.~\ref{fig:bias_dep_phase_amplitude}, we simulate topographic STM traces using a 1D model system  (\suppsec[SupSec:constant_current_topo]). In the simplest configuration corresponding to the Peierls reconstruction, we consider the contribution to the tunneling current of a single charge modulation (CM) and its associated gap centered on E$_F$ (Fig.~\ref{fig:two_gap_model}(a)-(c)). In this case, traces at the same polarity are always in phase, whereas traces at opposite polarities always show contrast inversion (or a 180$^{\circ}$ phase shift in the present harmonic model). The latter, often considered as an identifying hallmark of the CDW state~\cite{Spera2019}, clearly does not reproduce the data in Fig.~\ref{fig:bias_dep_phase_amplitude}(a). 

A single CM can only produce two sets of STM traces differing by contrast inversion in the vicinity of the gap. To generate a more complex bias dependence of the phase, we consider the possibility of a second CM whose associated gap opens in another band and away from $E_F$ (Fig.~\ref{fig:two_gap_model}(d)-(f)). If these two harmonic CMs are in phase (Fig.~\ref{fig:two_gap_model}(d)), the resulting STM traces are either in-phase or 180$^{\circ}$ out of phase ((Fig.~\ref{fig:two_gap_model}(e)), unable to reproduce the data in Fig.~\ref{fig:bias_dep_phase_amplitude}(a). To generate more structures in the bias dependence of the phase, we need to introduce a phase shift between the two CMs (Fig.~\ref{fig:two_gap_model}(g)-(i)). This leads to a phase which is no longer bi-modal, limited to two values differing by 180$^{\circ}$ as in (Figs.~\ref{fig:two_gap_model}(c)-(f)). It takes many different values (Fig.~\ref{fig:two_gap_model}(i)), where the precise bias dependence is defined by the magnitude of the two gaps, their position relative to $E_F$ and by the relative phase shift between the two CMs. The simulated STM topographic traces in Fig.~\ref{fig:two_gap_model}(b)(e) and (h) also reveal a pronounced bias dependent imaging amplitude with distinct line-shapes in the three model cases discussed above (\SF[fig:suppl_amplitude_in_models]). 

The broad parameter space of our 1D model makes it challenging to run a self converging fit to the data. Visually optimizing the size and position of the two gaps in Fig.~\ref{fig:two_gap_model}(i), we find a range of parameters (\suppsec[SupSec:two_gap])  simultaneously reproducing the experimental bias dependent CDW phase and amplitude data remarkably well (Fig.~\ref{fig:bias_dep_phase_amplitude}). As for the relative phase between the two CMs, it is chosen to minimize the Coulomb interaction of the CMs and to conform with the strong commensuration energy that locally locks them to the lattice. Reducing the Coulomb energy is obtained by introducing a phase shift between the two CMs, which can only be $\pm120^\circ$ ((Fig.~\ref{fig:two_gap_model}(g)) to satisfy the lock-in criterion with the lattice given the 3a$_0$ periodicity of the CDW.

\begin{figure}[htp]
\includegraphics{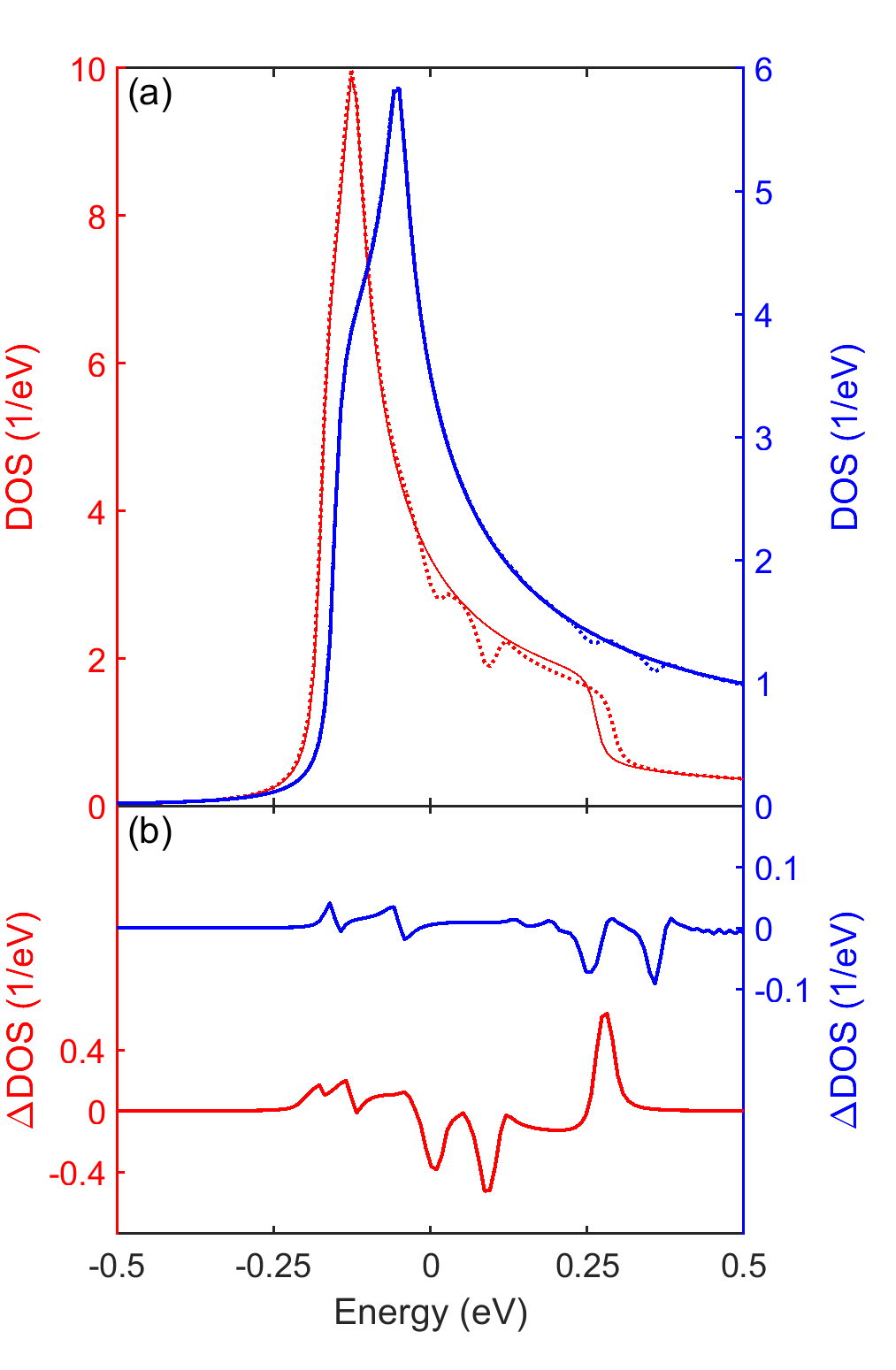}%
\caption{\label{fig:self_consistent} \textbf{Density of state of the two bands in the self-consistent calculations. } (a) The red (blue) indicates the band making up the inner (outer) pockets; solid/dashed indicates the ungapped/gapped bandstructure, where gaps are included self-consistently at the mean-field level. (b) Difference between the gapped and ungapped cases in (a), showing small DOS suppressions at energies away from E$_F$ on both bands in addition to the gap at E$_F$ on the inner band.}
\end{figure}

In the following, we turn to theoretically modelling 
multiple CDW gaps on different bands in NbSe$_2$. 
 We deploy self-consistent calculations to include the CDW gap within the random phase approximation on the two-dimensional two-band tight-binding fit to the NbSe$_2$ band structure constrained by ARPES \cite{Rahn2012}. The corresponding Fermi surface shown in Fig.~\ref{fig:struct_FS} consists of inner (red) and outer (blue) bands originating from symmetric and antisymmetric combinations of the Nb d$_{3z^2-r^2}$ orbitals. The model (see Methods) was previously shown to accurately reproduce the full range of experimental measurements on the charge-ordered state \cite{Flicker2015natcomm,Flicker2016prb}. The resulting DOS for the gapped and ungapped cases in each band are shown in Fig.~\ref{fig:self_consistent}(a). To emphasize the DOS reduction accompanying the CDW phase transition, we plot the difference between the gapped and ungapped DOS for each band in Fig.~\ref{fig:self_consistent}(b). 

Our theoretical modelling shows a clear gap on the inner band at $E_F$, consistent with the gap measured by ARPES around the $K$-point \cite{Rahn2012}. Interestingly, Fig.~\ref{fig:self_consistent}(b) reveals further DOS reductions, for example near 100 meV on the inner band and -50 meV on the outer band. These features are indicative of CDW gaps opening away from E$_F$ in addition to the (primary) gap at E$_F$, backing the simple model we propose to understand the bias dependence of the CDW appearance in STM images of NbSe$_2$. According to Fig.~\ref{fig:self_consistent}(b), there could even be more than two gaps. Consequently, we have included up to $n=8$ gaps to our 1D model. However, the agreement with the data is similar for $n=2$ and $n=3$ (\suppsec[SupSec:three_gap]), and we see no improvements adding more gaps. 

In summary, the remarkable match between the bias dependence of the CDW contrast in STM topography and the simple 1D model proposed here provides compelling evidence that the CDW in NbSe$_2$ is composed of at least two out-of-phase CMs on the inner and outer bands. While a 180$^\circ$ phase shift between these two CMs would minimise the Coulomb energy, the complex bias dependence of the CDW amplitude and phase observed by STM can only be reproduced when considering also the commensuration energy. This highlights the importance of the coupling of charge order to the lattice, which manifests in the formation of discommensurations~\cite{McMillan1976, Pasztor2019} and ultimately enables the observation of the multiband CMs uncovered here. The present study further highlights the power of topographic imaging to gain unique insight into detailed features of the CDW too faint to be detected accurately by tunneling spectroscopy. The formation of multiple modulations in response to new periodicities of a primary transition directly observed here, is extremely general and should in  principle be  present in all charge (and spin) density wave materials, and suggests new directions to explore in the physics of spatially modulated electronic orders.

\section{Methods}

\textbf{Crystal growth and STM measurements.} Single crystals of NbSe$_2$ were grown via iodine assisted chemical vapour transport and cleaved in-situ at room temperature. STM experiments were done in UHV (base pressure below $2\cdot10^{-10}$~mbar) using tips mechanically cut from a PtIr wire and conditioned in-situ on a clean Ag(111) single crystal. The bias voltage was applied to the sample. STM images were recorded in constant current mode with at least 64 pixel/nm resolution. Details of the CDW amplitude and phase fitting procedure can be found in Ref~\cite{Pasztor2019}. 

\bigskip

\textbf{Self-consistent calculations.} We modeled the CDW gap in the NbSe$_2$ bandstructure using the random phase approximation. Our model is based on a structured electron-phonon coupling dependent on the ingoing and outgoing electron momenta and the orbital character of the two bands scattered between; details are provided in Refs. \cite{Flicker2015natcomm, Flicker2016prb}. This model, which has as its only free parameter the overall magnitude of the electron-phonon coupling (constrained by $T_{CDW}$=33.5K), has previously been shown to agree well with the full range of experimental observations on the charge ordered state in NbSe$_2$. Using the Nambu Gor'kov method we solved for the CDW gap self-consistently at a set of high-symmetry points in the Brillouin zone, using the results to constrain a second tight-binding fit for the k-dependent CDW gap. We modeled the backfolding of the electronic bands using an approximate commensurate 3$\times$3 CDW geometry. A CDW gap at the Fermi level is observed in ARPES predominantly at points along the inner K-pocket, and our model reproduces this result. However, additional small gaps are also expected to open wherever band crossings are introduced by the backfolding (and where the CDW gap is nonzero), even at energies away from the Fermi level. This is the origin of the suppression of DOS at energies below E$_F$ seen in Fig.~\ref{fig:self_consistent}.

\section{Acknowledgments}
We acknowledge A. F. Morpurgo and I. Maggio-Aprile for inspiring discussions. We thank A. Guipet and G. Manfrini for their technical support in the STM laboratories. This project was supported by Div.2 (Grant No. 182652) and Sinergia (Grant No. 147607) of the Swiss National Science Foundation. F.~F.~acknowledges support from the Astor Junior Research Fellowship of New College, Oxford.  

\section{Author contributions}
C.R. designed the experiment. M.S. and A.S. took care of the STM experiments. \'A.P performed the data analysis. F.F. and J.v.W. performed the self-consistent calculations. \'A.P. did the one-dimensional model simulations. C.B. and E.G. synthesized the bulk crystals. \'A.P, A.S., F.F, J.v.W. and C.R. wrote the paper. All authors contributed to the scientific discussions and manuscript revisions.

\section{Competing interests}
The authors declare no competing interests.

\bibliography{multibandCDW_bibliography}

\end{document}